\documentstyle[preprint,aps,epsfig,floats]{revtex}
\tightenlines

\begin{document}

\title{Exact density profiles for the fully asymmetric exclusion
process \\
with discrete-time dynamics on semi-infinite chains \\}

\author{Jordan Brankov ${}^1$ and Nina Pesheva}
\address{Institute of Mechanics, Bulgarian Academy of Sciences, \\
Acad. G. Bonchev St. 4, 1113 Sofia, Bulgaria \\
e-mail: (1) brankov@bas.bg}
\maketitle
\date{submitted to the Phys. Rev. E}  

\date{\today}

\begin{abstract}
Exact density profiles in the steady state of the one-dimensional fully
asymmetric simple-exclusion process (FASEP) on a semi-infinite chain are
obtained in the case of forward-ordered sequential dynamics by taking the
thermodynamic limit in our recent exact results for a finite chain with
open boundaries. The corresponding results for sublattice parallel dynamics
follow from the relationship obtained by Rajewsky and Schreckenberg
[Physica A {\bf  245}, 139 (1997)], and for parallel dynamics from the
mapping found by Evans, Rajewsky and Speer [J. Stat. Phys. {\bf 95}, 45
(1999)]. Our analytical expressions involve Laplace-type integrals, rather
than complicated combinatorial expressions, which makes them convenient for
taking the limit of a semi-infinite chain, and for deriving the asymptotic
behavior of the density profiles at large distance from its end. By comparing
the asymptotic results appropriate for parallel update with those published
in the above cited paper by Evans, Rajewsky and Speer, we find complete
agreement except in two cases, in which we correct technical errors in the
final results given there.
\end{abstract}


\newpage

\section{Introduction}

We consider the current and density profiles in the steady state of the fully
asymmetric simple-exclusion process (FASEP) with open boundaries and different
discrete-time updates, namely, ordered sequential, sublattice parallel and
parallel. We remind the reader that the model describes a system of particles
on a chain, hopping with probability $p$ only to empty nearest-neighbor sites
to the right. Each of the $L$ sites of the chain can be either empty or
occupied by exactly one particle. Open boundary conditions mean that at each
time step (update of the whole chain) a particle is injected with probability
$\alpha$ at the left end of the chain ($i=1$), and removed with probability
$\beta$ at the right end ($i=L$). The definition of the model includes the
choice of the stochastic dynamics, i.e., the update scheme which specifies the
order in which the local hopping, injection and particle removal are
implemented. The case of random-sequential update is considered as a lattice
automaton realization of the corresponding continuous-time process. In the
general case it was solved first by using the recursion relations method
\cite{DDM}, \cite{SD}, and then
by means of the elegant matrix-product Ansatz (MPA) \cite{DEHP}. As it was
proven later \cite{KS}, the MPA is actually not an Ansatz, but an exact
representation of the stationary state of any one-dimensional system with
random-sequential dynamics involving nearest-neighbor hoppings and single-site 
boundary terms. The method of MPA was next successfully applied for solving
the following basic cases of true discrete-time dynamics: forward ordered
sequential ($\rightarrow$), backward ordered sequential ($\leftarrow$)
\cite{RSS}, \cite{HP}, and sublattice-parallel ($\mbox{s-}\parallel$) \cite{H},
\cite{RS}, which turned out to be closely related \cite{RSSS}. Thus, the
current has
the same value in all these cases, $J_L^{\rightarrow}=J_L^{\leftarrow}=
J_L^{\mbox{s-}\parallel}$; the local densities at site $i \in \{1,\dots ,L\}$
for the FASEP with forward, $\rho_L^{\rightarrow}(i)$, and backward,
$\rho_L^{\leftarrow}(i)$, ordered sequential updates are simply related to
each other, $\rho_L^{\rightarrow}(i)=\rho_L^{\leftarrow}(i)-J_L^{\rightarrow}$,
and to the local density $\rho_L^{\mbox{s-}\parallel}(i)$  for the
sublattice-parallel update \cite{RS}
\begin{equation}
\rho_L^{\mbox{s-}\parallel}(i)= \left\{ \begin{array}{ll}
\rho_L^{\rightarrow}(i),&\quad \mbox{$i$ odd} \\
\rho_L^{\leftarrow}(i),&\quad \mbox{$i$ even} .\end{array} \right.
\label{slor}
\end{equation}

In the cases of random-sequential, ordered sequential and sublattice parallel
updates, the matrix-product representation involves infinite dimensional
matrices which satisfy a particular quadratic algebra. Finally, the most
difficult case of fully parallel dynamics and open boundaries was solved by
Evans, Rajewsky, and Speer \cite{ERS}, by using site-oriented MPA with
matrices satisfying a quartic algebra. Moreover, these authors showed that the
current and local densities for the model with parallel update can be simply
mapped onto those for the previously solved discrete-time updates. An
independent, bond-oriented MPA solution for the stationary FASEP problem with
parallel dynamics was found by de Gier and Nienhuis \cite{GN}. In the case of
general values of the hopping probability $p$, they have presented explicit
expressions for the current and the discrete slope $t_L(i)=\rho_L(i+1) -
\rho_L(i)$ of density profile in all qualitatively different domains of
the parameter space.

In our recent paper \cite{BPV} we have derived, independently of \cite{ERS},
\cite{GN}, and by using a different method, exact expressions for the steady
state current $J_L^{\rightarrow}$ and the local density $\rho_L^{\rightarrow}
(i)$,  $i \in \{1,\dots ,L\}$, for the FASEP on a finite chain with
forward-ordered sequential dynamics and open
boundaries. These expressions involve integrals which at large $L$ and large
distance from the chain ends are of the Laplace type, hence, they are
convenient for asymptotic analysis. We point out that from the simple mapping
\cite{ERS} of the above quantities onto the current $J_L^{\parallel}$ and
local density $\rho_L^{\parallel}(i)$ for the model with parallel update,
\begin{equation}
J_L^{\parallel}={J_L^{\rightarrow}\over 1+J_L^{\rightarrow}} , \qquad
\rho_L^{\parallel}(i)={\rho_L^{\rightarrow}(i) +J_L^{\rightarrow}\over
1+J_L^{\rightarrow}} ,
\label{m1}
\end{equation}
exact new representations for $J_L^{\parallel}$ and $\rho_L^{\parallel}(i)$
follow, which do not contain complicated combinatorial expressions as those
obtained by Evans, Rajewsky, and Speer.
               
We shall present here exact expressions for the local density profiles
of the model on a semi-infinite chain with either left-hand (l) endpoint, 
\begin{equation}
\rho_{\infty, \rm l}^{\rightarrow}(i|\alpha,\beta) = 
\lim_{L \rightarrow \infty}\rho_L^{\rightarrow}(i|\alpha, \beta),
\label{leftp}
\end{equation}
or right-hand (r) endpoint,
\begin{equation}
\rho_{\infty, \rm r}^{\rightarrow}(j|\alpha,\beta)=\lim_{L\rightarrow \infty}
\rho_L^{\rightarrow}(L-j+1|\alpha,\beta).
\label{rightp}
\end{equation}
In the former case the limit $L\rightarrow \infty$ is 
taken at fixed $i=1,2,\dots$, which labels the sites of the chain from its
left end to the right. In the latter case the limit $L\rightarrow \infty$ is 
taken at fixed $j=1,2,\dots$, which labels the sites of the chain from its
right end to the left. Note that for fixed hopping probability $p$, and any
point $(\alpha, \beta)$ on the
phase diagram, see Fig.~1, the right-hand profile
$\rho_{\infty,\rm r}^{\rightarrow}(i|\alpha,\beta)$ is closely related to the
left-hand profile $\rho_{\infty, \rm l}^{\rightarrow}(i|\beta,\alpha)$ for
the point which is symmetrically positioned with respect
to the diagonal $\alpha = \beta$. Indeed, by taking the $L\rightarrow \infty$
limit in the symmetry relation for a finite chain \cite{RSSS},
$\rho_L^{\rightarrow}(i|\alpha, \beta)=1-J_L^{\rightarrow} -
\rho_L^{\rightarrow}(L-i+1|\beta, \alpha)$, one obtains that
\begin{equation}
\rho_{\infty,\rm l}^{\rightarrow}(i|\alpha,\beta)=1-
J_{\infty}^{\rightarrow} - \rho_{\infty,\rm r}^{\rightarrow}(i|\beta,\alpha).
\label{sym2}
\end{equation}
Therefore, it suffices to derive explicit expressions for the local densities
$\rho_{\infty,\rm l}^{\rightarrow}(i|\alpha,\beta)$
and $\rho_{\infty,\rm r}^{\rightarrow}(j|\alpha,\beta)$ for, say, $\alpha \leq
\beta$, i.e., in the maximum-current and low-density phases. Moreover, by
applying the $L \rightarrow \infty$ limit form of the mapping (\ref{m1}) with
$i$ fixed at a finite distance from one of the chain ends (e), 
\begin{equation}
J_{\infty}^{\parallel}={J_{\infty}^{\rightarrow}\over 1+
J_{\infty}^{\rightarrow}}, \qquad       
\rho_{\infty, \rm e}^{\parallel}(i)={\rho_{\infty, \rm e}^{\rightarrow}(i)
+ J_{\infty}^{\rightarrow}\over 1+J_{\infty}^{\rightarrow}} ,\quad
\mbox{(e = l,r)},
\label{m2}
\end{equation}
to the results for $\rho_{\infty, \rm e}^{\rightarrow}(i)$,
one obtains the corresponding expressions for
$\rho_{\infty, \rm e}^{\parallel}(i)$, $i=1,2,\dots$, e = l,r.
We emphasize that the resulting expressions for the density profiles on
semi-infinite chains are much simpler than those on finite chains,
and they again involve integrals which are of the Laplace type for large
distances from the chain end. Hence, their asymptotic behavior is readily
deduced. In order to make precise comparison with the asymptotic expansions
for the right-hand local density profiles
obtained in \cite{ERS}, one has to take into account that there, before
taking the limit $L\rightarrow \infty$, the sites are labeled from the right
end of the chain to the left by $n=L-i=0,1,2,\dots$, rather than by 
$j=L-i+1=1,2,\dots$, as in our definition (\ref{rightp}).   
By comparing the large $j$ asymptotic behavior of
$\rho_{\infty,\rm r}^{\parallel}(j)$, obtained via the mapping (\ref{m2}),
with the corresponding results of
Evans, Rajewsky and Speer \cite{ERS} for $\rho_n$ with $n=j-1$, we find
complete agreement in all but two cases, in which we correct technical
errors in the final asymptotic expressions (9.22) and (9.28) given there.
The expressions for the slope of the density profile derived in \cite{GN}
for general values of the hopping probability $p$ actually represent the
leading-order asymptotic form of $t_L(i)$ that follows from our
expressions for the local density $\rho_{L}(i)$ \cite{BPV} in the
semi-infinite chain limit $L \rightarrow \infty$ and in the leading-order
expansion in finite but large $L-i=n \gg 1$, see the results and comments
below. For recent general reviews on the MPA, ASEP and related systems we
refer the reader to \cite{DE}, \cite{D}, and \cite{CSS}. 

Concerning the notation, we make the following remarks. In contrast
to our previous work \cite{BPV}, here we shall adhere to the conventional
labeling of the different regions in the phase diagram, see Fig. 1. The mean
field line shown there is given by the equation $(1-\alpha)(1-\beta) = 1-p$, 
and the two other special lines are defined by
\begin{equation}
\alpha = \alpha_c \equiv 1-\sqrt{1-p}, \quad 
\beta = \beta_c \equiv 1-\sqrt{1-p}.
\label{acba}
\end{equation}          
The exact finite-chain results obtained in \cite{BPV} are conveniently
expressed in terms of the parametrs 
\begin{equation}
d = \sqrt{1-p},\quad a=d+d^{-1}, \quad \xi ={p-\alpha \over \alpha d},
\quad \eta = {p-\beta \over \beta d},
\label{dke}
\end{equation}
which will be used here too.

\section{Local density in the maximum-current phase}

The maximum-current phase occupies the region $\alpha_c < \alpha \leq 1$ and
$\beta_c < \beta \leq 1$, see region C in Fig. 1. In terms of the variables
(\ref{dke}) the above inequalities read: $-d\leq \xi<1$ and $-d \leq \eta <1$.

The final exact result obtained in \cite{BPV} for the local particle density
in the maximum-current phase is
\begin{equation}
\rho_L(i)={1\over 2}\left(1- J_L \right)+{d\over 2p Z_L}
[F_L(i)-(\xi -\eta)Z_{i-1}Z_{L-i}].
\label{70}
\end{equation}
Here $Z_n=Z_n(\xi,\eta)$ has the representation ($\xi \not= \eta$)
\begin{equation}
\label{51}
Z_n(\xi,\eta)= \left({d\over p}\right)^n \left[{\xi\over \xi-\eta}
I_n(\xi) + {\eta \over \eta -\xi}I_n(\eta)\right],
\end{equation}
which involves the integral
\begin{equation}
\label{51n}
I_n(\xi)= \frac{2}{\pi} \int_0^{\pi} {\rm d} \phi {(a+2
\cos\phi )^n \sin^2 \phi \over 1-2\xi\cos \phi +\xi^2} .
\end{equation}
The expression for $Z_n(\xi,\xi)$ follows by taking the limit
$\eta \rightarrow \xi$ in Eq. (\ref{51}). The current $J_L(\xi, \eta)$ is
given by 
\begin{equation}
\label{53}
J_L(\xi,\eta)=Z_{L-1}(\xi,\eta)/Z_L(\xi,\eta).
\end{equation}
The term $F_L(i)=F_L(i;\xi,\eta)$ in Eq. (\ref{70}) is an antisymmetric (with
respect to the center of the chain) function of the integer coordinate $i$,
$F_L(i;\xi,\eta)=-F_L(L-i+1;\xi, \eta)$,
defined for $1\leq i \leq [L/2]$ ($[x]$ denotes the integer part of $x$)
by the equation
\begin{equation}
\label{72}
F_L(i;\xi,\eta)= \left({d\over p}\right)^{L-1}(1-\xi \eta)
\sum_{n=0}^{L-2i}I_{L-i-n-1}(\xi)I_{i+n-1}(\eta).
\end{equation}

In order to take the limit $L\rightarrow \infty$ in Eq. (\ref{70}), we make
use of the large-$n$ asymptotic form of the Laplace-type integral (\ref{51n}): 
\begin{eqnarray}
I_n(\xi)&=& {(a+2)^{n+3/2}\over 2\sqrt{\pi}(1-\xi)^2}n^{-3/2}[1+O(n^{-1})],
\qquad (\xi \not= 1) \nonumber \\
I_n(1)&=& {(a+2)^{n+1/2}\over \sqrt{\pi}}n^{-1/2}[1+O(n^{-1})].
\label{54n}
\end{eqnarray}
By substituting the above expansion in Eq. (\ref{51}), we obtain that in the
maximum-current (m.c.) phase the leading asymptotic form of $Z_n(\xi,\eta)$
is ($\xi \not= 1$, $\eta \not=1$)
\begin{equation}
\label{54}
Z_n^{\rm m.c.}(\xi,\eta)=\frac{1-\xi \eta}{2\sqrt{\pi}(1-\xi)^2(1-\eta)^2}
\left({d\over p}\right)^n \frac{(a+2)^{n+3/2}}{n^{3/2}}[1+O(n^{-1})].
\end{equation}
Hence, the asymptotic form of the current reads
\begin{equation}
\label{55}
J_L^{\rm m.c.} =\frac{1-\sqrt{1-p}}{1+\sqrt{1-p}}[1+O(L^{-1})],
\end{equation}
independently of the parameters $\alpha$ and $\beta$. With the aid of
Eq. (\ref{54}), at fixed $i$ we obtain 
\begin{equation}
\lim_{L\rightarrow \infty}{d\over 2p}{Z_{i-1}Z_{L-i}\over Z_L} =
{\xi I_{i-1}(\xi) -\eta I_{i-1}(\eta)\over 2(\xi-\eta)(a+2)^i}.
\label{ZZ}
\end{equation}
Next, we split the sum over $n$ in Eq. (\ref{72}) into two parts, from
$0$ to $[L/2]-i$ and from $[L/2]-i+1$ to $L-2i$,
and replace the cofactor in the product $I_{L-i-n-1}(\xi)
I_{i+n-1}(\eta)$ which has large subscript as $L\rightarrow \infty$ 
by its asymptotic form (\ref{54n}). Thus, we prove the limit
\begin{eqnarray}
&&\lim_{L\rightarrow \infty}{d\over 2p}{F_L(i;\xi,\eta)\over Z_L(\xi,\eta)} =
\nonumber \\ &&{1\over 2}\sum_{n=0}^{\infty}\left[
{(1-\xi)^2I_{i+n-1}(\xi)\over (a+2)^{i+n+1}} +
{(1-\eta)^2I_{i+n-1}(\eta)\over (a+2)^{i+n+1}}\right].
\label{Flim}
\end{eqnarray}
Finally, by using the identity
\begin{equation}
{1\over 2}\sum_{n=0}^{\infty}{(1-\xi)^2I_{i+n-1}(\xi)\over (a+2)^{i+n+1}}=
{G_{i-1}-\xi I_{i-1}(\xi)\over (a+2)^i},
\label{iden}
\end{equation}
where
\begin{equation}
G_n = \frac{1}{\pi} \int_0^{\pi} {\rm d} \phi (a+2\cos\phi)^n (1+\cos \phi),
\label{G}
\end{equation}
and collecting the above results, we obtain an exact expression
for the local density profile of a semi-infinite chain with left-hand
endpoint in the form
\begin{equation}
\rho_{\infty,\rm l}^{\rightarrow}(i)={d\over 1+d} +
{1\over (a+2)^i}\left[G_{i-1}-\xi I_{i-1}(\xi)\right].
\label{profmc}
\end{equation}
The approach of the left-hand profile for finite chains, obtained by computer
simulations, to the limit
(\ref{profmc}) as the chain size increases is illustrated in Fig. 2.

The asymptotic behavior for $i\gg 1$ readily follows from
the expansions (\ref{54n}) and
\begin{equation}
G_{i-1}= {(a+2)^{i-1/2}\over \sqrt{\pi}}\left[i^{-1/2}-{a\over 16}i^{-3/2} +
O(i^{-5/2})\right].
\label{Gexp}
\end{equation}
Thus we obtain
\begin{eqnarray}
&&\rho_{\infty,\rm l}^{\rightarrow}(i)= {d\over 1+d} + {d^{1/2} \over
\sqrt{\pi}(1+d)}i^{-1/2} \nonumber \\&& -{1\over \sqrt{\pi}(1+d)}
\left[{2-p\over 16 d^{1/2}}+{\alpha d^{1/2}(p-\alpha)\over 2
(\alpha -\alpha_c)^2}\right]i^{-3/2} +O(i^{-5/2}).
\label{asmc}
\end{eqnarray}
Hence, by using the symmetry relation (\ref{sym2}) and the mapping (\ref{m2})
onto the case of parallel update, after taking into account the
correspondence 
$n=i-1$ with the expansion variable used in \cite{ERS}, we obtain
\begin{eqnarray}
&&\rho_n \equiv \rho_{\infty,\rm r}^{\parallel}(n+1)= {1\over 2} - {d^{1/2}
\over 2\sqrt{\pi}}n^{-1/2}  \nonumber \\&&
+\left[{2-p+8d\over 32 \sqrt{\pi}d^{1/2}}+{\beta d^{1/2}(p-\beta)\over 4
\sqrt{\pi}(\beta -\beta_c)^2}\right]n^{-3/2} +O(n^{-5/2}).
\label{cor1}
\end{eqnarray}
This result coincides with Eq. (9.22) of Ref. \cite{ERS} except for the sign
of the $O(n^{-1/2})$ term. The first two terms in the right-hand side of
Eq. (\ref{cor1}) generate Eq. (87) in \cite{GN}, provided the slope of the
density profile is considered there near the right-hand boundary of a
semi-infinite chain, i.e., when $L\rightarrow \infty$, $i/L\rightarrow 1$.

\section{Local density in the low-density phase}

The Low-density phase occupies the region $\alpha < \beta$ and $\alpha <
\alpha_c$, see region A=AI$\cup$AII in Fig. 1. In the case of a finite chain
of $L$ sites, up to exponentially small in $L$ corrections the current in the
low-density (l.d.) phase is \cite{BPV}
\begin{equation}
\label{64}
J_L^{\rm l.d.}(\xi,\eta) \simeq {p \over d(a +\xi +\xi^{-1})} =
\frac{\alpha(p-\alpha)}{p(1-\alpha)}.
\end{equation}

With respect to the asymptotic form of the density profile near the right-hand
end of the chain, one distinguishes two subregions, AI and AII.

{\bf Subregion AI}: $\alpha < \alpha_c$ and $\alpha < \beta < \beta_c$
($\xi> \eta >1$). As shown in \cite{BPV}, up to terms which are uniformly
in $i=1,\dots ,L$ exponentially small as $L\rightarrow \infty$, the local
density is
\begin{eqnarray}
\rho_L(i;\xi,\eta) \simeq {\alpha(1-p)\over p(1-\alpha)} +
{\eta -\eta^{-1} \over a+\xi +\xi^{-1}}\left({a+\eta +\eta^{-1} \over
 a+\xi +\xi^{-1}}\right)^{L-i}\nonumber \\
 -{\xi I_{L-i}(\xi)-\eta I_{L-i}(\eta) \over (a+\xi +\xi^{-1})^{L-i+1}}.
\label{y2}
\end{eqnarray}
Hence, in the limit $L\rightarrow \infty$ at fixed $i$, we obtain that
$\rho_{\infty,\rm l}^{\rightarrow}(i)$ is uniform, equal to
the bulk density given by the first term in the right-hand side of Eq.
(\ref{y2}). If the limit $L\rightarrow \infty$ is taken at fixed $j=L-i+1$, we
obtain the exact expression for the non-trivial density profile on a
semi-infinite chain with right-hand endpoint:
\begin{eqnarray}
\rho_{\infty,\rm r}^{\rightarrow}(j)={\alpha(1-p)\over p(1-\alpha)}
+{\eta -\eta^{-1} \over a+\eta +\eta^{-1}}\left({a+\eta +\eta^{-1} \over
 a+\xi +\xi^{-1}}\right)^j \nonumber \\
-{\xi I_{j-1}(\xi)-\eta I_{j-1}(\eta) \over (a+\xi +\xi^{-1})^j}.
\label{profld}
\end{eqnarray}

At large distance from the end of the chain, $j\gg 1$, the leading order
asymptotic form of $\rho_{\infty,\rm r}^{\rightarrow}(j)$
is given by the first two terms in the right-hand side of Eq. (\ref{profld}).
Upon mapping on the case of parallel dynamics, see Eq. (\ref{m2}), and taking
into account the labeling correspondence $j=n+1$, our result coincides with
Eq. (9.25) of \cite{ERS}. Then, the leading-order expression for the slope of
the density profile as $n\rightarrow \infty$ reproduces exactly Eq. (82) in
\cite{GN}.

{\bf Subregion AII}: $\alpha < \alpha_c$ and $\beta > \beta_c$ ($\xi >1$ and
$\eta <1$). As shown in \cite{BPV}, up to terms which are uniformly
in $i=1,\dots ,L$ exponentially small as $L\rightarrow \infty$, the local
density in this subregion is
\begin{equation}
\label{88}
\rho_L(i;\xi,\eta) \simeq {\alpha (1-p)\over p(1- \alpha)} -
{\xi I_{L-i}(\xi)- \eta I_{L-i}(\eta)\over (a+\xi +\xi^{-1})^{L-i+1}}.
\end{equation}
Hence, in the limit $L\rightarrow \infty$ at fixed $i$, we obtain that
$\rho_{\infty,\rm l}^{\rightarrow}(i)$ is again uniform, equal to
the bulk density. When the limit $L\rightarrow \infty$ is taken at fixed
$j=L-i+1$, we obtain the exact expression for the non-trivial density profile
on a semi-infinite chain with right-hand endpoint:
\begin{equation}
\rho_{\infty,\rm r}^{\rightarrow}(j)={\alpha(1-p)\over p(1-\alpha)}
-{\xi I_{j-1}(\xi)-\eta I_{j-1}(\eta) \over (a+\xi +\xi^{-1})^j}.
\label{profld2}
\end{equation}

At large distance from the right-hand endpoint, $j\gg 1$, the leading order
asymptotic form of $\rho_{\infty,\rm r}^{\rightarrow}(j)$
in subregion AII is
\begin{eqnarray}
&&\rho_{\infty,\rm r}^{\rightarrow}(j)={\alpha(1-p)\over p(1-\alpha)}
\nonumber \\&& -{\sqrt{a+2}\over 2\sqrt{\pi}}\left[{\xi \over (1-\xi)^2}
-{\eta \over (1-\eta)^2}\right]\left({a+2 \over a+\xi +\xi^{-1}}
\right)^j j^{-3/2}+O(j^{-5/2}).
\label{profld3}
\end{eqnarray}
A comparison of the right-hand density profile for a finite chain,
Eq. (\ref{88}), and its leading-order asymptotic form on a semi-infinite chain,
Eq. (\ref{profld3}), with the result of computer simulations for $L=300$ is
given in Fig. 3.

Upon mapping on the case of parallel dynamics, see Eq. (\ref{m2}),
expressing the result in terms of the original parameters $\alpha$ and $\beta$,
and taking into account the labeling correspondence $j=n+1$, we obtain
\begin{eqnarray}
&&\rho_{\infty,\rm r}^{\parallel}(n)={\alpha(1-\alpha)\over p-\alpha^2} -
{\alpha (p-\alpha)\over 2\sqrt{\pi}(p-\alpha^2)}{d^{1/2}\over 1-d}
\nonumber \\&&\times \left[{\alpha (p-\alpha)\over (\alpha-\alpha_c)^2}
-{\beta (p-\beta)\over (\beta-\beta_c)^2}\right]
\left[{\alpha (p-\alpha)\over (1-\alpha)(1-d)^2}\right]^n n^{-3/2}
\nonumber \\&& +O\left(n^{-5/2}\left[{\alpha (p-\alpha)\over (1-\alpha)
(1-d)^2}\right] ^n \right).
\label{profld4}
\end{eqnarray}
This is the correct version of Eq. (9.28) in \cite{ERS}; the error there
occured on passing to the second equality in Eq. (9.27). For the slope of the
density profile the first two terms in the right-hand side of
Eq.~(\ref{profld4}) yield Eq.~(84) in \cite{GN} as a leading-order
expression when $n \gg 1$. 

{\bf The boundary between subregions AI and AII}: $\alpha <\alpha_c$ and
$\beta = \beta_c$ ($\xi >1$ and $\eta =1$). The exact expression for the
density profile on a semi-infinite chain with right-hand endpoint is given
by Eq. (\ref{profld2}) at $\eta =1$. Its large-distance asymptotic form is
redily obtained with the aid of expansions (\ref{54n}):
\begin{eqnarray}
&&\rho_{\infty,\rm r}^{\rightarrow}(j)={\alpha(1-p)\over p(1-\alpha)}
\nonumber \\&& +{1\over \sqrt{\pi}\sqrt{a+2}}\left({a+2 \over a+\xi +
\xi^{-1}}\right)^j j^{-1/2}+O(j^{-3/2}).
\label{profld5}
\end{eqnarray}
Upon mapping on the case of parallel dynamics with the aid of Eq. (\ref{m2}),
we recover exactly Eq. (9.29) in \cite{ERS}. Then, for the leading-order
asymptotic form of the slope of the density profile as $n\rightarrow \infty$
one obtains Eq. (83) in \cite{GN}.

\section{Discussion}

Summarizing, we have calculated exactly the local density profiles in all
the phases of the FASEP on semi-infinite chains with open either left-hand
(particle injection) or right-hand (particle removal) boundary.
The obtained expressions are much simpler than the finite-size ones, and they
readily admit large-distance asymptotic analysis.

The following general features of the local density profiles on semi-infinite
chains turn out to be common to all the basic updates:

(i) The profile in the maximum-current phase depends on the boundary
condition at the finite endpoint of the chain, and {\em does not} depend
on the boundary condition at infinity. The leading-order approach to the
bulk density with the distance $i$ from the chain end is of the order
$O(i^{-1/2})$ and does not depend on the boundary conditions at all.

(ii) The local density is uniform in the low-density phase of a chain with
left-hand endpoint, and in the high-density phase of a chain with right-hand
endpoint.

(iii) The local density profiles in the low-density phase of a chain with
right-hand endpoint, and in the high-density phase of a chain with left-hand
endpoint, depend on the boundary conditions {\em both} at the finite endpoint
and at infinity. This fact is reminiscent of the dependence of equilibrium
states with spontaneously broken symmetry on boundary conditions at infinity.
However, it is the continuity of the current that keeps here the
information about the boundary condition at the chain end that goes to
infinity in the limit $L\rightarrow \infty$.

(iv) The analytic form of the local density profile on a semi-infinite chain
with right-hand (left-hand) endpoint changes on passing from subregion AI (BI)
to subregion AII (BII) within the low-density (high-density) phase, compare
Eqs. (\ref{profld}) and (\ref{profld2}). The leading-order large-distance
asymptotic approach to the bulk density is also qualitatively different: in AI
(BI) the approach is purely exponential, with inverse correlation length 
$\lambda^{-1} = |\lambda_{\xi}^{-1} - \lambda_{\eta}^{-1}|$, while in AII (BII)
the approach is exponential, with power-law prefactor $j^{-3/2}$
and inverse correlation length $\lambda_{\xi}^{-1}$ ($\lambda_{\eta}^{-1}$),
where
\begin{equation}
\lambda_{\xi}^{-1} = \ln \left({a+\xi+\xi^{-1} \over a+2}\right).
\label{cordt}
\end{equation} 
On the borderline between these two subregions the leading-order asymptotic
approach is exponential, with power-law prefactor $j^{-1/2}$ and inverse
correlation length given by $\lambda_{\xi}^{-1}$ ($\lambda_{\eta}^{-1}$).
On crossing the mean-field line in subregion AII (BII), the analytic
form of the profile does not change, only its bending near the right-hand
(left-hand) end changes from downward (upward) above that line to upward
(downward) below it, see Eq. (\ref{profld3}).

(v) The correlation length depends on the type of update only: for all the
true discrete-time updates $\lambda_{\xi}$ is given by Eq. (\ref{cordt}), 
and for the random sequential update, see Eq. (78) in \cite{DEHP},
\begin{equation}
\lambda_{\alpha}^{-1} = \ln \left({1 \over 4\alpha (1-\alpha)}\right).
\label{corct}
\end{equation} 

We point out that from the mapping (\ref{m1}) of our exact results for the
finite-chain current $J_L^{\rightarrow}$ and local density
$\rho_L^{\rightarrow}(i)$ \cite{BPV} onto the case of parallel update,
exact new representations for $J_L^{\parallel}$ and $\rho_L^{\parallel}(i)$
follow, which, in contrast to those found in \cite{ERS}, are convenient for
asymptotic analysis. Finally, by comparing the so obtained asymptotic results
to those published in \cite{ERS}, we have corrected technical errors in two of
the expressions given there. The leading-order large-distance expressions that 
follow for the slope of the density profile near the right-hand endpoint of a
semi-infinite chain reproduce the corresponding results obtained in \cite{GN}.

\newpage

\begin{figure}
\epsfxsize=15cm
\epsfysize=15cm
\epsfig{file=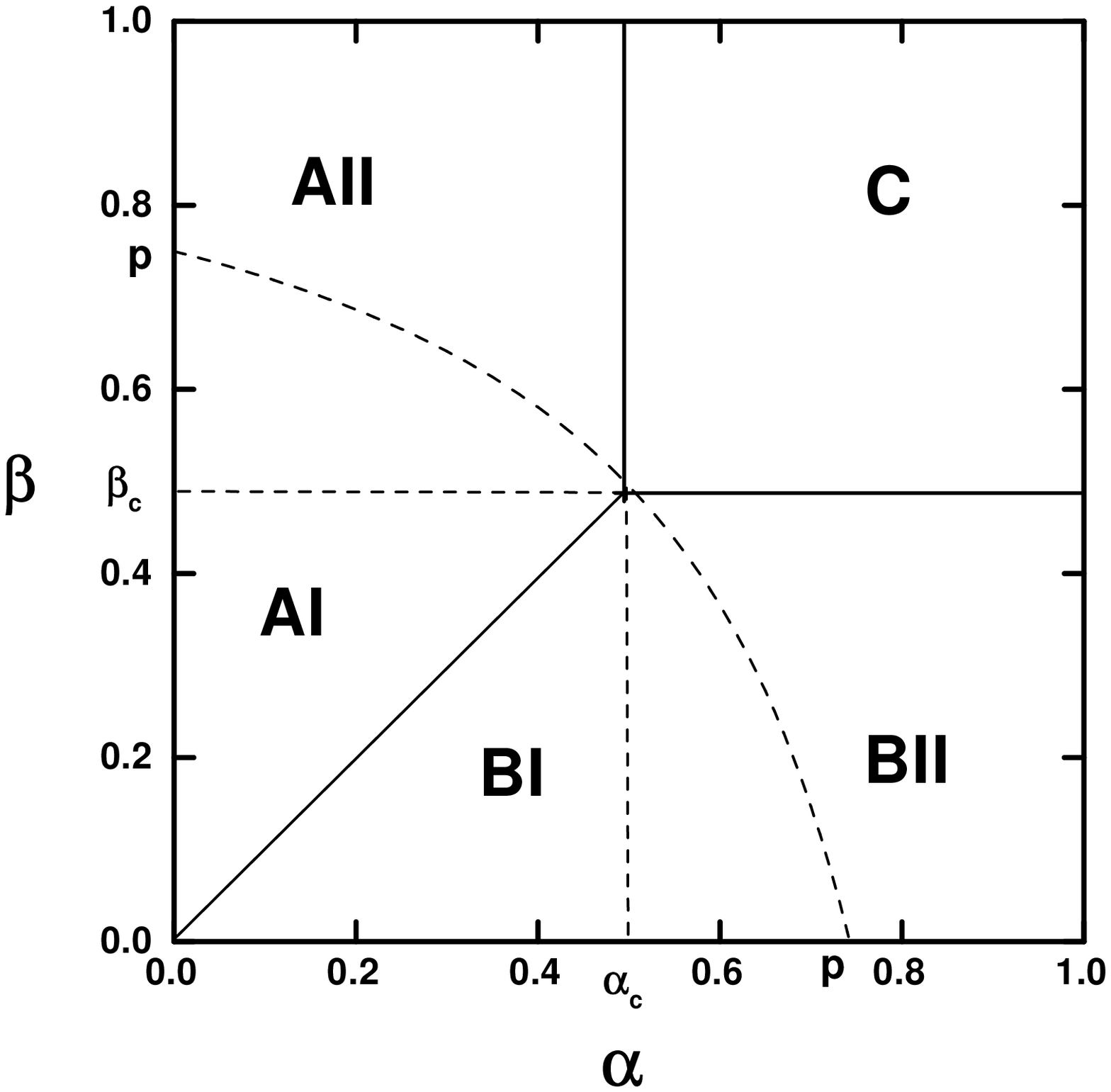}
\caption{The phase diagram in the plane of the injection and removal
probabilities $\alpha$ and $\beta$ (see the text) for hopping probability
$p$=0.75. The maximum-current phase occupies region {\bf C}. Region
{\bf A}$=${\bf AI}$\cup${\bf AII} corresponds to the low-density phase, and
region {\bf B}$=${\bf BI}$\cup${\bf BII} to the high-density phase.
Subregions {\bf AI} ({\bf BI}) and {\bf AII} ({\bf BII}) are distinguished by
the different analytic form of the density profile. The boundary between
them, $\beta = \beta_c$, $0\leq \alpha \leq \alpha_c$
($\alpha = \alpha_c$, $0\leq \beta \leq \beta_c$), is shown by dashed
segment of a straight line. The solid line $\alpha = \beta$ between subregions
{\bf AI} and {\bf BI} is the coexistence line of the low- and high-density
phases. The curved dashed line is the mean-field line $(1-\alpha)(1-\beta) =
1-p$.}
\label{fig:1}
\end{figure}

\begin{figure}
\epsfxsize=16cm
\epsfysize=13cm
\epsfig{file=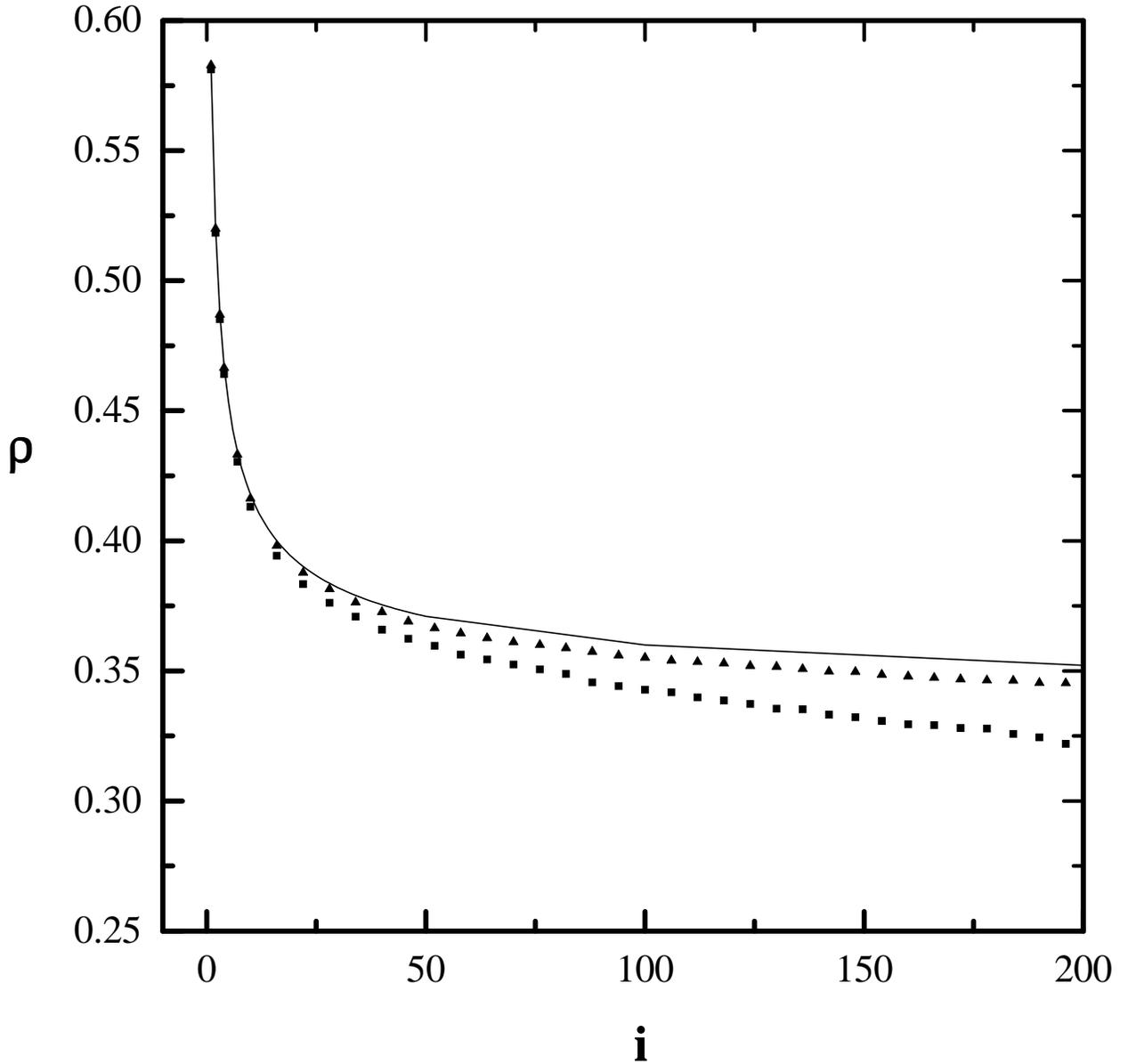}
\caption{The approach of the left-hand local number-density profile $\rho$
versus the site number $i$ in the
maximum-current phase ($\alpha = \beta =0.8$, $p=0.75$), obtained by computer
simulations for finite chains of length $L$: solid squares - $L=300$, solid
triangles - $L=1000$ (data averaged over 300 runs of $2^{16}$ Monte Carlo
steps per site), to the limit given by Eq. (21) - solid line.}
\label{fig:2}
\end{figure}

\begin{figure}
\epsfxsize=16cm
\epsfysize=13cm
\epsfig{file=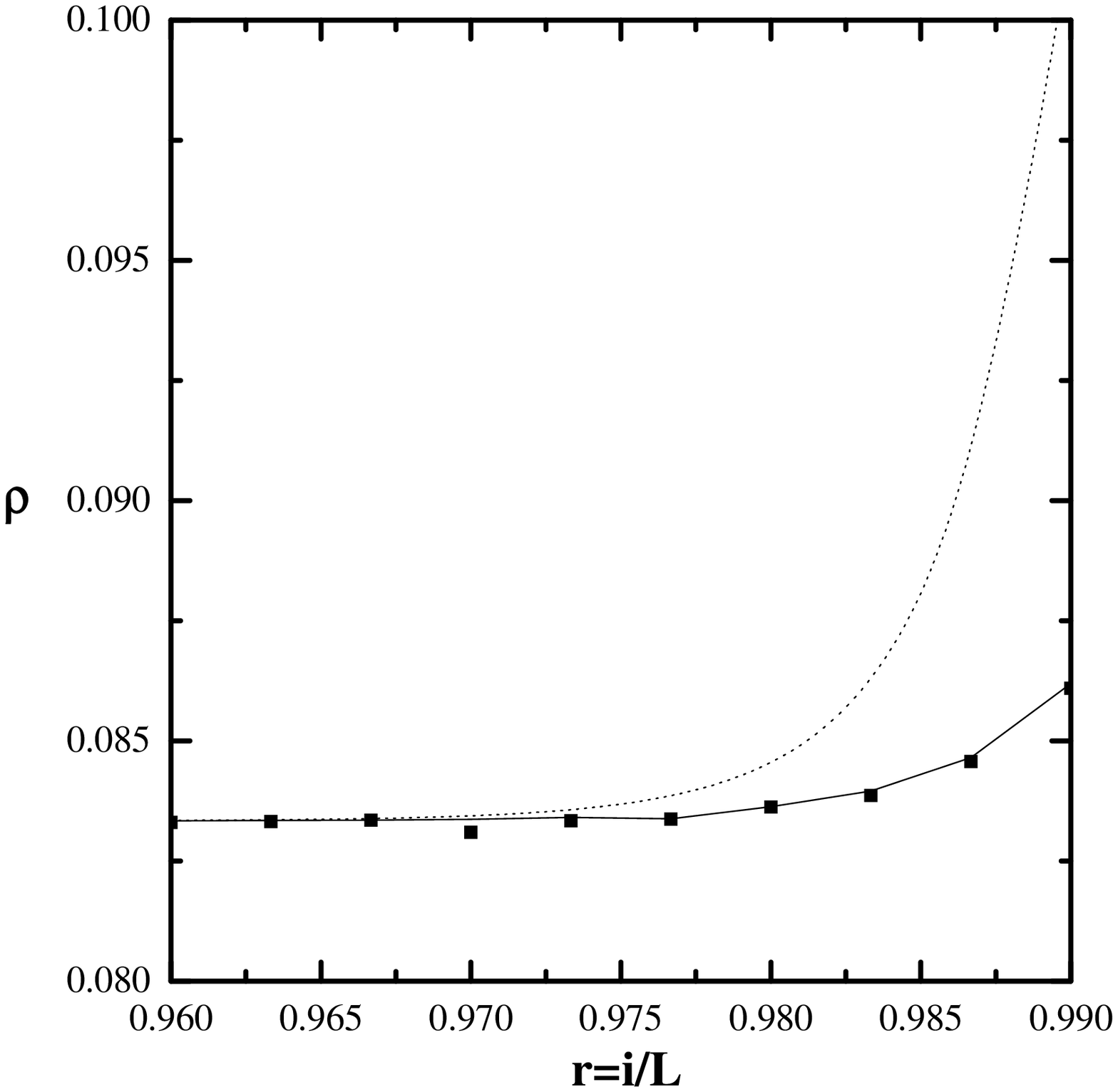}
\caption{The behavior of the right-hand number-density profile $\rho$
versus the scaled distance $r=i/L$ in subregion
{\bf AII} ($\alpha = 0.2$, $\beta =0.6$, $p=0.75$): the solid squares present
the results of computer
simulations for a finite chain of $L=300$ sites,
the solid line corresponds to Eq. (29) for  $L=300$, and the dotted
line - to the leading-order asymptotic form of the profile for a semi-infinite
chain given by Eq. (30).}
\label{fig:3}
\end{figure}


\begin{thebibliography}{99}

\bibitem{DDM} D. Derrida, E. Domany, and D. Mukamel,
J. Stat. Phys. {\bf 69}, 667 (1992). 

\bibitem{SD}
G. Sch\"{u}tz and E. Domany, J. Stat. Phys. {\bf 72}, 277 (1993).

\bibitem{DEHP}
B. Derrida, M. R. Evans, V. Hakim, and V. Pasquier, J. Phys. A 
{\bf 26}, 1493 (1993).

\bibitem{KS}
K. Krebs and S. Sandow, J. Phys. A {\bf 30}, 3165 (1997).

\bibitem{RSS}
N. Rajewsky, A. Schadschneider, and M. Schreckenberg, J. Phys. A 
{\bf 29}, L305 (1996).

\bibitem{HP}
A. Honecker and I. Peschel, J. Stat. Phys. {\bf 88}, 319 (1997).

\bibitem{H}
H. Hinrichsen, J. Phys. A {\bf 29}, 3659 (1996)

\bibitem{RS}
N. Rajewsky and M. Schreckenberg, Physica A {\bf  245}, 139 (1997).

\bibitem{RSSS} N. Rajewsky, L. Santen, A. Schadschneider, and M. Schreckenberg,
J. Stat. Phys. {\bf 92}, 151 (1998). 

\bibitem{ERS}
M. R. Evans, N. Rajewsky, and E. R. Speer,
J. Stat. Phys. {\bf 95}, 45 (1999).

\bibitem{GN}
J. de Gier and B. Nienhuis, Phys. Rev. E {\bf 59}, 4899 (1999).

\bibitem{BPV}
J. Brankov, N. Pesheva, and N. Valkov, Phys. Rev. E {\bf 61}, 2300 (2000).

\bibitem{DE}
B. Derrida and M. Evans, in: {\it Nonequilibrium Statistical Mechanics in One
Dimension}, ed. V. Privman (Cambridge University Press, Cambridge, 1997).

\bibitem{D}
B. Derrida, Phys. Rep. {\bf 301}, 65 (1998).

\bibitem{CSS}
D. Chowdhury, L. Santen, and A. Schadschneider, Phys. Rep. {\bf 329}, 199
(2000).

\end{thebibliography}
\end{document}